\documentclass[a4paper,11pt]{article}
\usepackage[utf8]{inputenc}
\usepackage{amsmath}
\usepackage{amsfonts}
\usepackage{amssymb}
\usepackage{amsthm}
\usepackage{caption}
\usepackage{multirow}
\usepackage{fontenc}
\usepackage{graphicx}
\usepackage{color}
\usepackage{hyperref}
\usepackage[symbol]{footmisc}
\usepackage{subcaption}
\usepackage{cleveref}
\usepackage{gensymb}
\usepackage[total={6in, 9.5in}]{geometry}

\date{}
                
\def\title#1{\begin{center} {\LARGE #1 \vspace{0.3cm}} \end{center}}
\def\author#1{\begin{center}{ \large #1} \end{center}}
\def\affil#1{\begin{center}{ \it #1} \end{center}}

\begin{document}


\title{The hadronic beamline of the \textsc{ENUBET} neutrino beam} 

\author{C.~Delogu, on behalf of the ENUBET collaboration\footnote[1]{F.~Acerbi, A.~Berra, M.~Bonesini, A.~Branca, C.~Brizzolari, G.~Brunetti, M.~Calviani, S.~Capelli, S.~Carturan, M.G.~Catanesi, S.~Cecchini, N.~Charitonidis, F.~Cindolo, G.~Collazuol, E.~Conti, F.~Dal~Corso, G.~De~Rosa, A.~Falcone, A.~Gola, C.~Jollet, V.~Kain, B.~Klicek, Y.~Kudenko, M.~Laveder, A.~Longhin, L.~Ludovici, E.~Lutsenko, L.~Magaletti, G.~Mandrioli, A.~Margotti, V.~Mascagna, N.~Mauri, L.~Meazza, A.~Meregaglia, M.~Mezzetto, M.~Nessi, A.~Paoloni, M.~Pari, E.~Parozzi, L.~Pasqualini, G.~Paternoster, L.~Patrizii, M.~Pozzato, M.~Prest, F.~Pupilli, E.~Radicioni, C.~Riccio, A.C.~Ruggeri, C.~Scian, G.~Sirri, M.~Stipcevic, M.~Tenti, F.~Terranova, M.~Torti, E.~Vallazza, F.~Velotti, M.~Vesco, L.~Votano}} 

\affil{INFN Sezione di Padova, via Marzolo 8, Padova, Italy \\
Department of Physics, Università di Padova, via Marzolo 8, Padova, Italy }

 

\begin{abstract}
The knowledge of initial flux, energy and flavour of current neutrino beams is currently the main limitation for a precise measurement of neutrino cross sections. The ENUBET ERC project, part of the CERN Neutrino Platform as NP06/ENUBET, is studying a facility based on a narrow band beam capable of constraining the neutrino fluxes normalization through the monitoring of the associated charged leptons in an instrumented decay tunnel (tagger). Furthermore, in narrow-band beams, the transverse position of the neutrino interaction at the detector can be exploited to determine a priori with significant precision the neutrino energy spectrum without relying on the final state reconstruction.

A key element of the project is the design and optimization of the hadronic beamline. It requires an efficient focusing and collimation over short distances in order to minimize the fraction of secondary mesons decaying before the tunnel entrance and to reduce the background on the tunnel walls coming from particles different from the large angle decay products.



This poster will present progress on the studies of the proton extraction schemes. It will also show a realistic implementation and simulation of the beamline, both in a configuration with a single dipole magnet for charge and momentum selection and in a recently studied one with a double dipole providing a larger bendind angle, with benecial effects on the beam halo background and on the untagged neutrino component at the far detector.
\end{abstract}

\vspace*{\stretch{1}}

\begin{center}
{\LARGE{Presented at}}\\
\vspace{0.3cm}
\Large{NuPhys2019: Prospects in Neutrino Physics\\
Cavendish Conference Centre, London, 16--18 December 2019}
\end{center}
 
\clearpage
 
\section{ENUBET - Enhanced NeUtrino BEams from kaon Tagging}
In the near future neutrino physics will require measurements of absolute neutrino cross sections at the GeV scale with 1\% precision. Modern cross section experiments are reaching the intrinsic limitations of conventional neutrino beams: $\nu_e$ and $\nu_{\mu}$ fluxes are inferred by a full simulation of meson production and transport from the target down to the beam dump and are validated by external data and, hence, neutrino fluxes are affected by significant uncertainties, of the order of 5-10\%.

The goal of the ENUBET project\footnote[1]{This project has received funding from the European Research Council (ERC) under the European Unions Horizon 2020 research and innovation programme (Grant Agreement 681647).} is to study a facility where the production of electron neutrinos from $K_{e3}$ decays can be monitored on a single particle level by instrumenting the decay region in a narrow band neutrino beam. 
Positrons produced in association with electron neutrinos in $K_{e3}$ ($K^{+} \rightarrow e^{+} \pi^{0} \nu_{e}$) decay are tagged and monitored in an instrumented decay tunnel \cite{Longhin:2014yta}.
The beamline is designed to enhance the $\nu_e$ components from $K_{e3}$ and to suppress the $\nu_e$ contaminations from muon decay.
This would allow to measure $\nu_e$ cross sections with a precision improved by about one order of magnitude, compared to present results.

Two main pillars are required for the success of the project:
\begin{itemize}
\item the design and construction of a detector capable of performing positron identification in a $\nu$ beam decay tunnel at single particle level \cite{posterAntonio}; 
\item the layout of the $\pi/K$ focusing and transport system with suitable proton extraction schemes to keep the rate of particles in the tunnel at a level sustainable for the tunnel instrumentation.
\end{itemize}
Secondary particles produced by proton interactions in the target are focused and transported to the decay pipe. Non-interacting protons are stopped in a beam dump. Off-momentum particles reaching the decay tunnel are mostly low energy particles coming from interactions in other beamline components and muons that cross absorbers and collimators.

The optimization of the ENUBET beamline is performed taking into account different requirements \cite{Acerbi:2645532, Acerbi:2019qiv}:
\begin{itemize}
	\item Maximize the number of $K^+$ in the momentum range of interest at tunnel entrance.
	\item Minimize the total length of the transferline ($\sim$20 m) to reduce kaon decay losses before the entrance of the decay tunnel.
	\item Produce a small beam size: non decaying particles should exit the decay pipe without hitting the tagger inner surface.
	\item Keep under control the level of background transported to the tunnel, which affects the signal-to-noise ratio of the positron selection.
	\item Use of conventional magnet field and apertures (normal-conducting devices, with an aperture below 40 cm).
\end{itemize}
The optics is currently optimized for a reference hadron beam with a momentum of 8.5~GeV/c and a momentum bite of 10$\%$.

The secondary particles production from the interaction of primary protons with the target are simulated with FLUKA. The optimization of the components of the beamline (dipoles and quadrupoles) is performed with TRANSPORT. A full simulation of particle transport and interaction is performed with G4beamline. Assessment of the doses is addressed using FLUKA. 

\section{``Static'' focusing beamline}
The first proposed focusing system, the static one, consists of a quadrupole triplet placed before the bending magnet, downstream the target. This configuration allows to perform the focusing using DC operated devices, instead of pulsed magnetic horns, offering several advantages in terms of costs and technical implementation.

\begin{figure}[htbp]
	\centering
	\begin{subfigure}[t]{0.65\textwidth}
		\includegraphics[width=\textwidth]{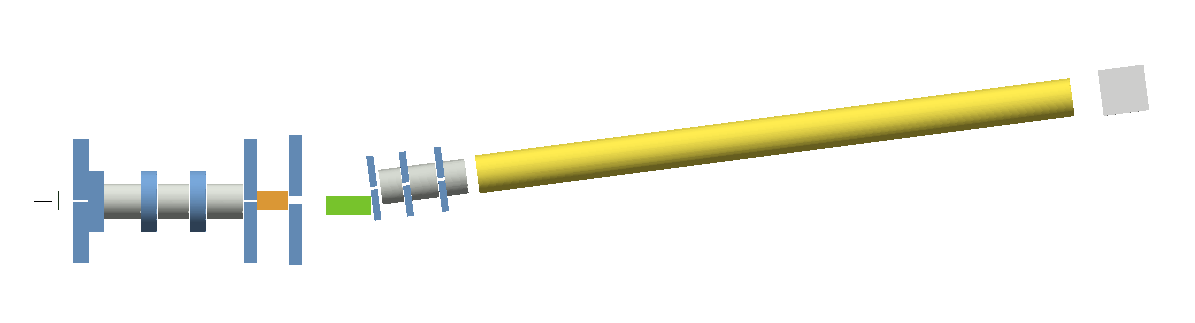}
        \caption{Single dipole beamline}
		\label{11}
	\end{subfigure}
	\hfill
	\begin{subfigure}[t]{0.3\textwidth}
		\includegraphics[width=\textwidth]{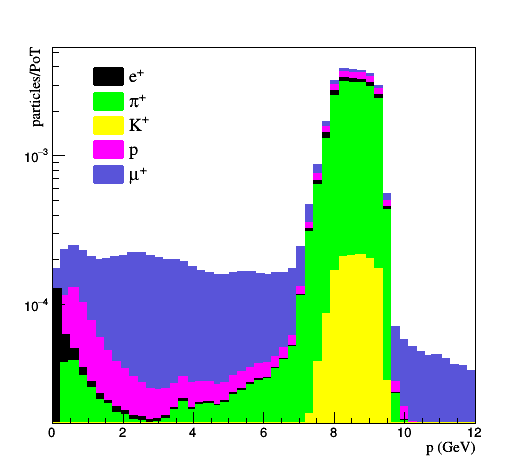}
		\caption{Particle budget at tagger entrance}
		\label{12}
	\end{subfigure}
	\caption{Static focusing system.}
	\label{tlr2}
\end{figure}

The reference option consists in a quadrupole triplet followed by a dipole, that provides a 7.4$\degree$ bending angle, and by another quadrupole triplet (Fig.\ref{11}).
One advantage with respect to an horn-based line is that there are no intrinsic time limits for proton extractions, up to several seconds. The single resonant slow extraction (2 s) is less challenging than a horn-based beamline. Moreover, the particle rate at the tunnel instrumentation could be reduced, reducing pile-up effects.
Fig.\ref{12} shows the expected beam composition at the detector entrance.

\begin{figure}[h!]
\centering
	\includegraphics[width=.7\textwidth]{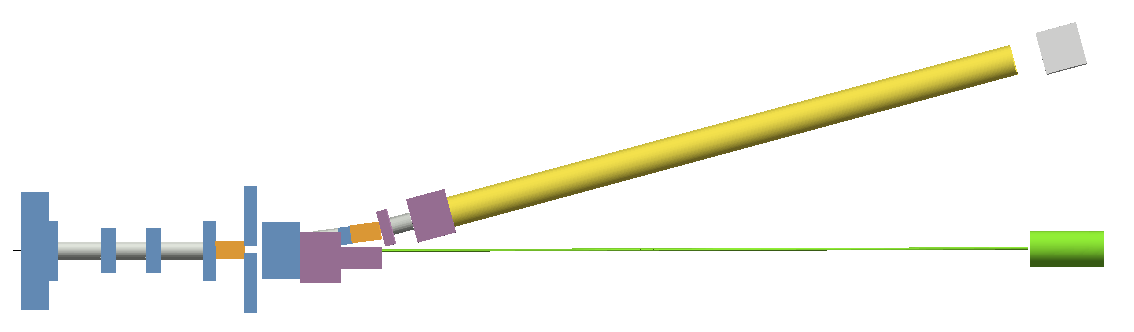}
	\caption{Schematics of the ENUBET double dipole beamline.}
	\label{tlr5}
\end{figure}

A new design is the double dipole configuration (Fig.\ref{tlr5}). The bending angle is larger, 15.2$\degree$, and the length of this transferline is greater than the single dipole option. Significant advantages of this design are a reduction of the beam halo background (in particular from muons) and of the untagged neutrino component at the far detector (neutrinos produced in the straight section of the transfer line have a lower probability to reach the detector).

Moreover, since the instantaneous rate of particles hitting the decay tunnel walls is reduced, compared with the horn option, a neutrino interaction in the detector could be linked with the observation of its associated lepton in the decay tunnel. This could lead to a facility where the neutrino is uniquely associated with the other decay particle, a so called ``tagged neutrino beam''. 

\section{Horn-based beamline - ``burst slow extraction''}
\begin{figure}[h!]
	\centering
	\includegraphics[width=8.cm]{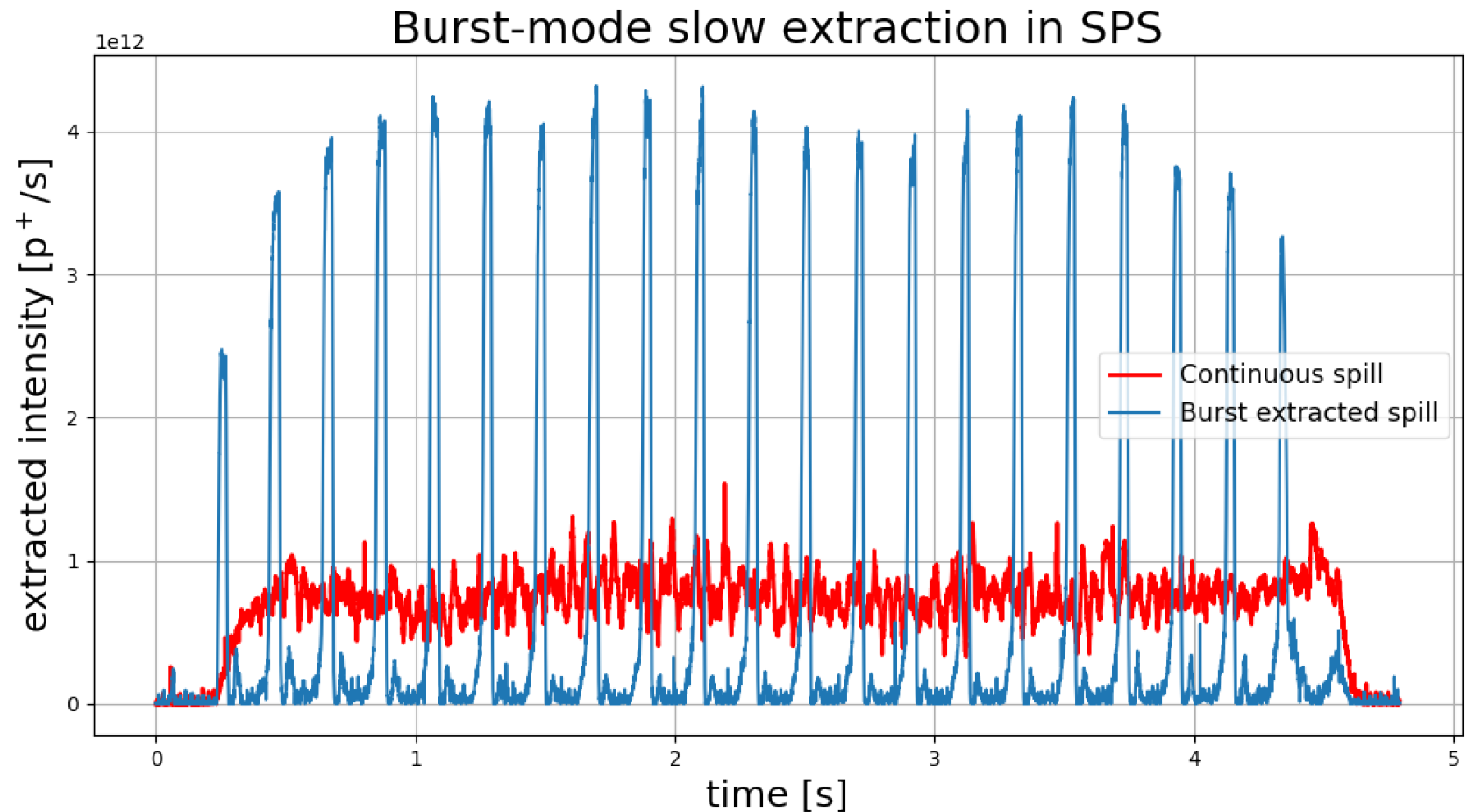}
	\caption{Burst slow extraction over a SPS spill \cite{Pari:2019bhy}.}
	\label{sps}
\end{figure}

This option features a magnetic horn, placed between the target and the quadrupoles. It has to be pulsed with large currents (180 kA) for 2-10 ms and cycled at 10 Hz during the accelerator flat-top.
Studies concerning the proton extraction scheme (``burst slow extraction'') to synchronise a few ms proton extraction with current pulsing are on-going at the CERN-SPS (Fig.\ref{sps}).
It is more challenging than the single resonant slow extraction over O(s) times (as in the static option), but higher yields can be achieved at the tunnel entrance using a magnetic horn instead of the quadrupole triplet (Tab.\ref{focusingschemesrates}).


\begin{table}[ht!]
	\centering
	\begin{tabular}{|cccc|}
		\hline
		Focusing system & $\pi^+/POT$ (10$^{-3})$ & $K^+/POT$ (10$^{-3}$) & Extraction length \\
		\hline 
		Horn-based & 77 & 7.9 & 2-10~ms\\ 
		Static & 19 & 1.4 & 2~s\\ 
		\hline
	\end{tabular} 	
	\caption{Expected rates of $\pi^+$ and $K^+$ in 6.5$\div$10.5~GeV range at the decay tunnel entrance for the two possible focusing schemes.}
	\label{focusingschemesrates}
\end{table}


\end{document}